\documentclass[12pt,preprint]{aastex}
\usepackage{url}
\usepackage{natbib}
\usepackage{aas_macros}
\shorttitle{Using Realistic MHD Simulations for Modeling and Interpretation of Quiet-Sun Observations with HMI}

\title{Using Realistic MHD Simulations for Modeling and Interpretation of Quiet-Sun Observations with the Solar Dynamics Observatory Helioseismic and Magnetic Imager}

\author{I.~N. Kitiashvili$^{1,2}$, S.~Couvidat$^2$, A.~Lagg$^3$}
\affil{$^1$NASA Ames Research Center, Moffett Field, Mountain View, CA 94035, USA}
\affil{$^2$Stanford University, Stanford, CA 94305, USA}
\affil{$^3$Max Planck Institute for Solar System Research,  G\"ottingen, 37077, Germany}

\begin{document}
\begin{abstract}
The solar atmosphere is extremely dynamic, and many important phenomena develop on small scales that are unresolved in observations with the Helioseismic and Magnetic Imager (HMI) instrument on the Solar Dynamics Observatory (SDO). For correct calibration and interpretation of the observations, it is very important to investigate the effects of small-scale structures and dynamics on the HMI observables, such as Doppler shift, continuum intensity, spectral line depth, and width. We use 3D radiative hydrodynamics simulations of the upper turbulent convective layer and the atmosphere of the Sun, and a spectro-polarimetric radiative transfer code to study observational characteristics of the Fe~I~6173~\AA~line observed by HMI in quiet-Sun regions. We use the modeling results to investigate the sensitivity of the line Doppler shift to plasma velocity, and also sensitivities of the line parameters to plasma temperature and density, and determine effective line formation heights for observations of solar regions located at different distances from the disc center. These estimates are important for the interpretation of helioseismology measurements. In addition, we consider various center-to-limb effects, such as convective blue-shift, variations of helioseismic travel-times, and the `concave' Sun effect, and show that the simulations can qualitatively reproduce the observed phenomena, indicating that these effects are related to a complex interaction of the solar dynamics and radiative transfer.
\end{abstract}


\section{Introduction}
The {\it Helioseismic and Magnetic Imager} (HMI) instrument onboard the {\it Solar Dynamics Observatory} \citep{Schou2012} satellite provides a unique opportunity to observe the solar surface dynamics over many spatial and temporal scales and to investigate various phenomena in quiet-Sun and active regions \citep{Scherrer2012,Zhao2012a}. The HMI instrument continuously obtains near real-time images of the full solar disc in the Fe~I ($\lambda_0=6173.3$~\AA) absorption line. With the HMI front camera, measurements of the line profile are taken for two circular polarizations at six wavelength positions located symmetrically with respect to the reference wavelength, $\lambda_0$. The resulting filtergrams are converted into the line-of-sight (LOS) observables: continuum intensity, Doppler velocity, and LOS magnetic field \citep{Couvidat2012}. A detailed investigation of the line properties is important for data calibration, interpretation, and explanation of various observational effects. In this paper, we consider the HMI observables for quiet-Sun conditions without a magnetic field.

Generally, there are two approaches to investigate the relationship between the instrumental observables and the corresponding physical properties of the Sun: 1) use 1D or 3D models of the solar atmosphere and a radiative transfer code to calculate variations of a synthetic line profile and the observables to infinitely small perturbations of the physical properties (so-called, the instrumental response function); 2) use 3D dynamical models of the solar atmosphere from realistic numerical simulations and a radiative transfer code to model the observables and then calculate the Pearson correlation coefficients with the actual physical parameters, as a function of height. The first approach represents a classical astronomical problem \citep[for a recent formulation and references see ][]{RuizCobo1994}; it was applied to filtergram-type observations of solar spectral lines, and specifically to the SOHO Michelson Doppler Imager (MDI)  data by \citep{Wachter2008}. The second approach \citep[e.g.][]{Fleck2011} became  available only recently with the development of realistic numerical simulations. A primary goal of both approaches is to estimate an effective height of the line formation in the solar atmosphere, and its variations for different conditions and center-to-limb positions. The advantage of the first approach that in the 1D case it allows certain freedom in choosing the  reference atmospheric models. For instance, this allows to investigate various semi-empirical sunspot models \citep[see e.g.][]{Wachter2008}. The second approach while providing a more realistic relationship between the observables and the actual solar conditions is restricted the available numerical simulations which represent a substantial computational effort. In both cases, the sensitivity function is spread over a range of heights; however, the inferences obtained by these two approaches have a different meaning. In the first approach, the calculated sensitivity function depends only on the optical line formation properties (absorption coefficient and source function). In the second approach, the atmospheric properties are dynamically coupled through the radiative MHD equations, and therefore, the correlation coefficients depend not only on the optical depth but also on the physical conditions in a broader range of heights. This means, for example, that because the atmospheric dynamics is driven by deeper  turbulent motions  the instrumental observables are also sensitive to these motions. In this paper, we present the results obtained by both these approaches for the atmospheric models from our numerical simulations. However, dynamical effects (such as, the observed convective blue shift, helioseismic time shifts, and p-mode line asymmetry) can be correctly interpreted only using the second approach. Currently, because of the computational limitations the realistic 3D simulations can be performed only for relatively small computational domains, and do not allow to  a full quantitative analysis of helioseismology measurements. However, they provide an important insight into the origin of the observed effects, and show the importance of this approach for understanding the dynamical coupling in the solar atmosphere, and directions for further investigations.

For this investigation we use 3D radiative hydrodynamics simulations of solar convection \citep{Kitiashvili2013a,Kitiashvili2013b}, and the radiation transfer code  SPINOR/STOPRO \citep{Solanki1992,Frutiger2000}. We investigate the sensitivity of the HMI observables to atmospheric properties as functions of height, and study the origin and properties of  the center-to-limb effects, such as the convective blue shift \citep[e.g.][]{Hastings1873,Halm1907,Evershed1931,Adam1976} and the artifact of a `shrinking' or 'concave' Sun observed in helioseismic travel-time shifts \citep{Duvall2009,Zhao2012}. We also calculate variations of the effective line formation height for various distance from the disc center \citep[e.g.][]{Carlsson2004,Norton2006,Fleck2011} and determine effects of the HMI point spread function. We plan to study the role of background magnetic fields and the magnetic field observables in a separate paper.

\section{Modeling of HMI observables} \label{s:1}
The 3D radiative hydrodynamics simulations are obtained using the `SolarBox' code developed at NASA/Ames. These simulation results are converted into synthetic line profiles with the SPINOR/STOPRO code \citep{Solanki1992,Frutiger2000}; then the Doppler shift, continuum intensity, and line depth are calculated using a simplified version of the HMI pipeline codes \citep{Couvidat2012a}. By comparing the Doppler-shift velocity with the actual plasma velocity in the simplified atmosphere, we determine the effective line formation height. Similarly, we investigate the sensitivity of the continuum intensity and line depth to temperature and density variations at different heights in the solar atmosphere.

\subsection{Numerical model} \label{s:solarbox}
Accurate modeling of the solar turbulent convection is critical for a correct representation of the small-scale dynamics, which is a key element of energy exchange between different scales. In this study we use a 3D radiative hydrodynamics code developed for modeling the top layers of the convective zone and the low atmosphere from first principles \citep[e.g.][]{Jacoutot2008,Kitiashvili2013b,Kitiashvili2013a}. The code is based on a Large Eddy Simulation (LES) formulation for compressible flow and includes a fully coupled radiation solver, assuming Local Thermodynamic Equilibrium (LTE) conditions. The radiative transfer between the fluid elements is calculated using a 3D multi-spectral-bin long-characteristics method. Typically, we use 4 spectral bins. The ray-tracing transport is calculated using the \cite{Feautrier1964} method with 14 rays, and taking into account ionization and excitation of all abundant species. Our 3D numerical model includes time and space dependent physical quantities which are converted in to 1D dynamic atmospheric models and used as input for the SPINOR code \citep{Frutiger2000}.
 
The computational domain of $6.4\times6.4\times6.2$~Mm includes a 1-Mm layer of the low atmosphere. The grid-size is 12.5~km in the horizontal directions; a variable grid-spacing of similar size is set in the vertical direction. The lateral boundary conditions are periodic. The top boundary is open to mass, momentum, and energy fluxes, and also to the radiation flux. The bottom boundary is open only for radiation, and simulates the energy input from the interior of the Sun. 
The simulation is initialized from a standard solar model of the interior structure and the lower atmosphere \citep{Christensen-Dalsgaard1996}, to which are added small random velocity perturbations ($\pm 1-2$ cm/s) and run until a relaxed statistically stationary state of solar convection is reached. 

\subsection{Synthetic line profiles}\label{s:stokes}
\begin{figure}
\centerline{\includegraphics[scale=1]{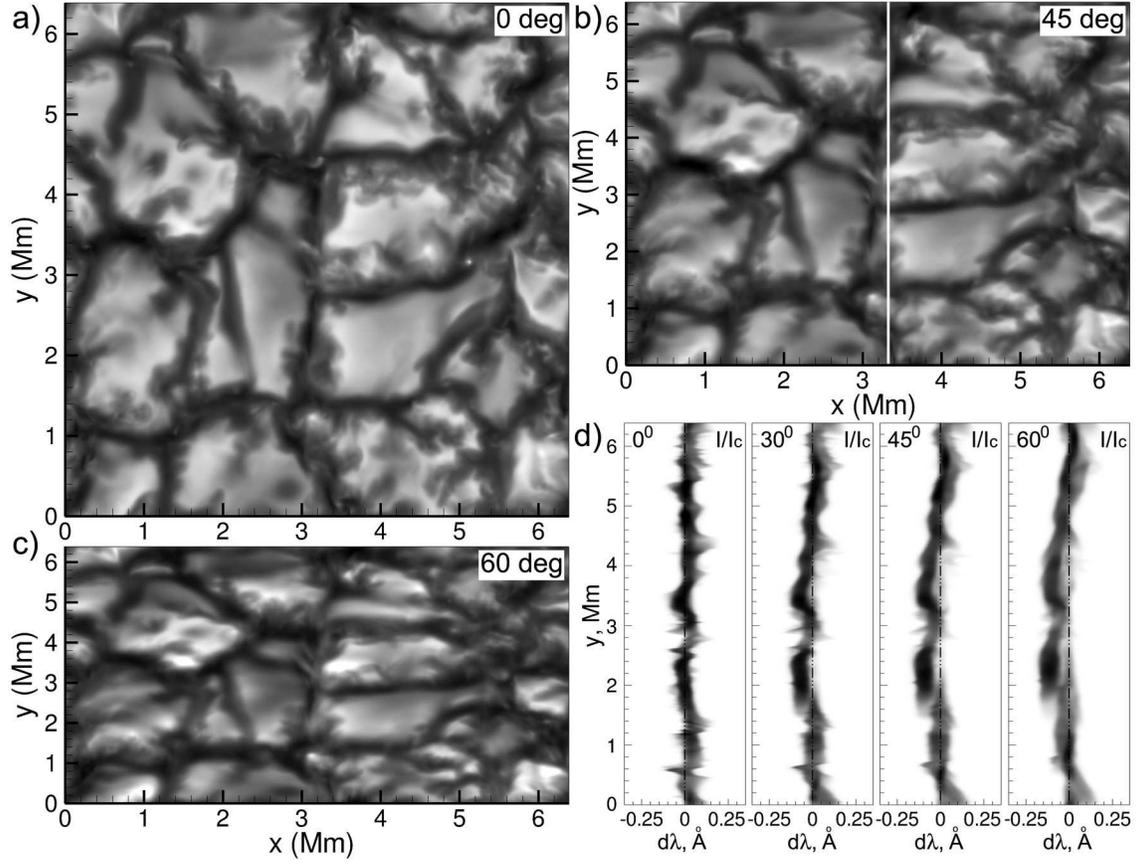}}
 \caption{Continuum intensity of the Fe~I, $\lambda_0 = 6173.3$~\AA~line calculated for different angular distances from the disc center: $0^0$ (panel $a$), $45^0$ ($b$) and $60^0$ (panel $c$). Panel~$d$ shows variations of the line profiles along a vertical slit (indicated by white line in panel $b$).} \label{Icont}
\end{figure}

We use data-cubes of the hydrodynamic simulations, saved with a time cadence of 25~sec, as a dynamical model of the solar atmosphere for input into the  radiative transfer SPINOR/STOPRO code \citep{Solanki1992,Frutiger2000,vanNoort2012}. This code is developed for solving the forward radiation transfer problem in the LTE approximation (STOPRO package) and can also perform inversions of observational spectropolarimetric data. Using a recent MPI-parallelized version of SPINOR, we calculate profiles for the HMI line Fe~I,~$\lambda_0=6173.3$~\AA~as a function of wavelength for different angular distances from the disc center: $0^0$, $30^0$, $45^0$ and $60^0$. In Figure~\ref{Icont}, we illustrate the calculated continuum intensity data for three snapshots at 0, 45, and 60 degrees from the disc center (the heliographic distances correspond to the viewing line-of-sight angle with respect to the radial direction). At the 60-degree viewing angle the granules appear as `hills' on the solar surface. Panel ($d$) shows the synthetic line profiles for the same locations along a slit indicated by the vertical line in panel  ($b$). The simulated spectrograms reveal systematic changes of the line properties: Doppler shift, line width, and depth with distance from disc center. Evidently, the line width and the systematic line shifts become larger for larger distances. The resulting mean profile of the Fe~I~6173~\AA~line is in good agreement with the FTS solar atlas\footnote{\it available at ftp://ftp.hs.uni-hamburg.de/pub/outgoing/FTS-Atlas} \citep{Brault1987}. The maximum deviation from the observed profile is about 2\% for iron abundance of 7.43~dex and much less than 1\% for the abundance of 7.5~dex (Fig.~\ref{line_prof}).

\subsection{Synthetic HMI observables}\label{s:HMI}
The observables (Doppler shift, continuum intensity, and line depth) are calculated by applying a simplified version of the HMI pipeline procedure \citep{Couvidat2012}  to the 10-hour time-series of the simulated line profiles. These synthetic observables are obtained for different line-of-sight angles, and also for the full numerical resolution (12.5~km) and the HMI resolution ($\sim 315$~km/pixel) calculated by convolving the simulated observables with the theoretical HMI point spread function (PSF) \citep{Wachter2012}. 

Figure~\ref{Doppler} shows the Doppler shift snapshots for the same moment of time as in Figure~\ref{Icont} for the disc center and $60^0$ distance from the center, for both the computational grid resolution (panels $a$ and $c$) and the HMI resolution (panels $b$ and $d$). The full-resolution Dopplergrams reveal complicated small-scale turbulent dynamics on subgranular scales (e.g. vortex tubes, shearing flows, see Fig.~\ref{Doppler}~$a$, $c$) which are not resolved by HMI. The HMI PSF filters out the subgranular scales and also makes the intergranular lanes wider than they are in the realistic simulations (Fig.~\ref{Doppler}). 

The sensitivity of the spectral line (also called the instrumental response function; Ruiz Cobo \& del Toro Iniesta 1994) to variations of physical properties of the solar atmosphere is an important quantity, which requires a special detailed study. Previously, the instrumental response function for the filtergram type of measurements for a hydrostatic solar 1D model was studied by \cite{Wachter2008}. 
 In Figure~\ref{RF} we present the results of calculations of the response function for the HMI Dopplergrams to line-of-sight velocity perturbations and HMI Continuum Intensity to temperature perturbations. The calculations are performed for the mean model of the solar atmosphere using the radiative transfer code SPINOR/STOPRO and the same procedure for the response function as described by Ruiz \& del Toro Iniesta (1994) and Wachter (2008). Our results are qualitatively in good agreement with Wachter(2008), but show that the maximum of the Dopplergram response function for the HMI instrument is about 125 km compared to 175 km for the MDI instrument (see Fig. 3 of Wachter, 2008).

\section{Center-to-limb effects}
It is well known that line-of-sight observations of solar regions at different distances from the disc center are often accompanied by systematic deviations related to foreshortening due to the spherical geometry and projection effects. In addition, because the inclined light path through the solar atmosphere is longer for the larger distance, the physical conditions of line formation depend on the distance from the disc center, affecting the line formation height. These deviations are often treated through various empirical corrections, such as removal of the limb darkening, shift of the Fraunhofer lines and others \citep[e.g.][]{Hastings1873,Halm1907,Evershed1931,Adam1976}. With the development of modern observational and data analysis methods new details and phenomena related to the center-to-limb variations are found \citep[e.g.][]{Stenflo1997,BellotRubio2009,Duvall2009,Couvidat2012a,Bonet2012,Zhao2012}. The understanding and interpretation of these effects require sophisticated numerical modeling. Here we present some initial results of such modeling.
 
\subsection{Variations of the line formation height }\label{s:height}
Estimation of spectral line formation heights is critical for understanding the local physical conditions reflected in the observables. Previous studies showed that solar granulation, when viewed away from the disc center, represents  a `hilly'-like surface (Fig.~\ref{Icont} $c$) because the line formation height varies from point to point  \citep{Carlsson2004,Fleck2011}. In the current study we are interested in the mean formation height of the HMI Doppler shift observed in line 6173~\AA,  and, therefore, following the approach of \cite{Fleck2011}, we calculate the Pearson correlation coefficients of the simulated plasma velocities along the line-of-sight and the corresponding Doppler shift for different distances from the disc center. Figure~\ref{corr-coeff} shows the correlation coefficients for the cases with the full numerical resolution (black and red curves) and for the HMI-resolution (pink and blue curves). The maxima of the correlation coefficients correspond to the effective line formation height for the Doppler-shift data. Thin curves illustrate the vertical profile of the correlation coefficients for different moments of time (with 5~min cadence). Thick red and blue curves correspond to the mean vertical profile of the correlation coefficient, averaged over 8~hours.

Figure~\ref{corr-coeff} clearly shows strong differences between the line formation heights estimated from the data with the numerical and HMI-like resolutions. The dependences of the formation height on the distance from the disc center is shown in Figure~\ref{form}. The difference between the full and degraded resolutions, however, decreases for the regions closer to the solar limb. The smoothing of the projected velocity field along the line-of-sight makes the correlation coefficients more  similar for large distances from the disc center (Fig.~\ref{corr-coeff}). Also, the difference in the estimated line formation heights between the full and degraded resolutions decreases with the distance (Fig.~\ref{form}). 

Similarly, we calculate the Pearson correlation coefficients between the HMI continuum intensity, and temperature and density in the simulated atmosphere. The results, presented in Figure~\ref{corr-coeff_Ic-T}, show that the continuum intensity is sensitive to the temperature gradient in the range of heights of 0 -- 200~km, and to the density at 0 -- 400~km. We did not find significant differences in the correlation coefficients for the continuum intensity and the line depth.

\subsection{Convective blue shift effect}\label{s:blue}
Among the well-known manifestations of the center-to-limb variations are correlated variations of the granular intensity and a systematic Doppler shift, known as the convective blue-shift effect \citep{Adam1976}. The observed mixture of blue and red shifts on the solar surface reflects the dynamics of convective patterns, such as granules, which are characterized by upflow motions in granules (blue shift), and downflows in the intergranular lanes (red shifted). The dark intergranular lanes cover substantially smaller area than the bright granules; therefore the contribution of granules in the mean Doppler shift is greater, and this causes the predominant blue-shift of the absorption lines which are formed near the photosphere. A decrease of the line blue-shift and possible transition from blue to red shifts at larger angular distances reflects the decreasing intensity contrast and decreasing contribution of the upflows. To reproduce this effect, we plot spatially-averaged variations of the Doppler velocity as a function of time (Fig.~\ref{velocity_angles_vs_time}~$a$, $c$) for different distances from the disc center (or line-of-sight viewing angles with respect to the radial direction), from $0^0$ (disc center, blue curve) to $60^0$ (near the limb, red curve). Velocity fluctuations which correspond to the 5-minute oscillations, are observed at all simulated angular distances; however, the velocity profiles show systematic positive shifts with increasing angular distance (Figs.~\ref{velocity_angles_vs_time}~$a$, $b$). Comparison of the time-averaged Doppler-shift velocities of the simulated quiet-Sun area for different angular distances shows a systematic `blue' shift of about 50~m/s for the disc center and `red' shift of about 260~m/s at $60^0$ (Fig.~\ref{velocity_angles_vs_time}~$b$). After removing the systematic blue shift, Doppler velocities (Fig.~\ref{velocity_angles_vs_time}~$c$) show a decrease of amplitude of the 5-min oscillations for larger angular distances, for which the contribution of the radial flows decreases. 

\subsection{Helioseismic travel times and the `shrinking Sun' effect}\label{s:shrinking}
Observations of solar oscillations are used by helioseismology and allows us to infer properties of the subsurface dynamics of the Sun \citep[see for reviews,][]{Christensen-Dalsgaard2004,Kosovichev2011}. A challenging task is to measure the meridional circulation in the solar convection zone. This requires very accurate measurements of travel times or phase shifts covering a wide range of distances from the disc center to the solar limb. Recent measurements of  \cite{Zhao2012} revealed unexpected systematic center-to-limb effects in acoustic travel times, which affected the determination of deep meridional flows. The problem first arose when \cite{Duvall2009} found that the oscillation signal from a limb region arrives about 2~sec faster, compared to the disc center, than was expected from theory. This apparent artifact was dubbed as a  `{\it shrinking}' (or `{\it concave}') {\it Sun} effect. Later, helioseismic high-resolution measurements of the near-full solar disc also demonstrated that the systematic center-to-disc variations are different for different variables: $\sim 2$~s for the HMI Dopplergrams, and $\sim 10$~s for the HMI continuum intensity data \citep{Zhao2012}. These apparent artifacts could be due to instrumental and data processing problems or have a physical origin due to changes in the line formation conditions. It was suggested that the travel time shifts can be due to changes in line formation height \cite{Zhao2012}; the asymmetrical nature of solar granulation was proposed by \cite{Baldner2012}; also, radiative heat transfer effects were discussed by \cite{Duvall2009}.

To investigate the presence of such variations in the numerical simulations, we calculate cross-correlation functions between the mean Doppler velocities of the same region, but observed at different angles. The results (Fig.~\ref{velocity_angles_vs_time}$d$) clearly show a decreasing  time-delay signal with increasing angular distance, up to $\sim 2$~s for $60^0$. For the continuum intensity this time delay increases to $\sim 3$~sec (Fig.~\ref{cross-corr}). The existence of such deviations in the numerical model favors the physical origin of the effect, and also indicates that it can be due to the changes in the line formation height, as shown in Figures~\ref{corr-coeff} and~\ref{form}. 

\subsection{Solar oscillations}\label{s:oscil}
Stochastically excited by turbulent convection, 5-minute oscillations (Fig.~\ref{Doppler}$a$, $c$) carry an important information about subsurface dynamics of the Sun. Helioseismology applications observe oscillations at horizontal wavelengths larger than our box domain. Therefore, we consider properties of horizontally averaged oscillations. Observations of these oscillations for different angular distances reflect line-of-sight projection effects of the radial oscillations, variations of the line formation height, and contributions from convective noise. Figure~\ref{spectrum_velocity_angles} shows the power spectral density of the simulated Doppler-shift signal for different line-of-sight distances from the disk center. The peaks in the power spectra correspond to the radial modes reflected at the box bottom boundary. As observed on the Sun, the power of the oscillation modes decreases with the distance, and mode lines display the line asymmetry also corresponding to the observed asymmetry \citep{Nigam1998}. At the lowest frequencies the power spectrum is dominated by convective flows that one predominantly horizontal. This results in power increase with the distance from the disk center at these frequencies.

\section{Discussion and conclusion}

Realistic 3D radiative hydrodynamic simulations allow us to investigate important links between solar observations and the physical and dynamical properties of solar conditions, taking into account spectral line formation conditions, line-of-sight effects, and instrumental spatial and spectral resolutions. For detailed comparison of observations and numerical models it is important to model the observational and data analysis procedures and calculate the `observables' using the simulation data. Such comparison is necessary for validation of the models and also for calibration and interpretation of observations. The full-disc high-resolution observations by the  {\it Helioseismic and Magnetic Imager} (HMI) onboard the {\it Solar Dynamics Observatory} \citep{Schou2012} have opened new opportunities to investigate solar dynamics and probe deeper flows in the solar interior, but they also require careful calibration and accounting for the center-to-limb effects that are potentially significant for high-precision measurement. 

To investigate systematic deviations in the observational data for different angular distances from the disc center, we use radiative hydrodynamic simulations of quiet-Sun regions without magnetic field. The simulated data sets are used as an evolving in time model of the solar  atmosphere, and these were used to derive the HMI line Fe~I,~$\lambda_0=6173$~ {\AA} properties by employing the SPINOR radiation transfer LTE code \citep{Solanki1992,Frutiger2000,vanNoort2012}. 

\begin{figure}
\centerline{\includegraphics[scale=0.5]{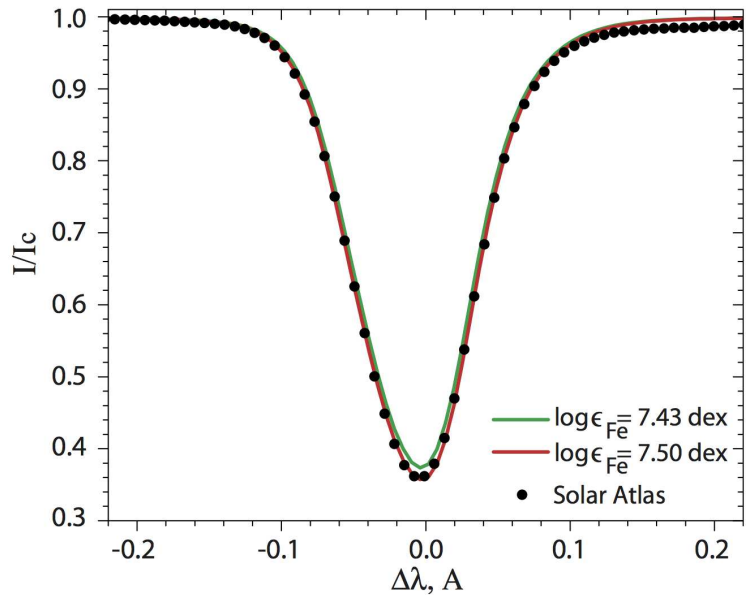}}
 \caption{Comparison of the mean profile of the Fe~I, $\lambda_0 = 6173.3$~\AA~line at the solar disc center calculated for the different iron abundances and FTS solar atlas \citep{Brault1987}.} \label{line_prof}
\end{figure}

\begin{figure}
\centerline{\includegraphics[scale=1]{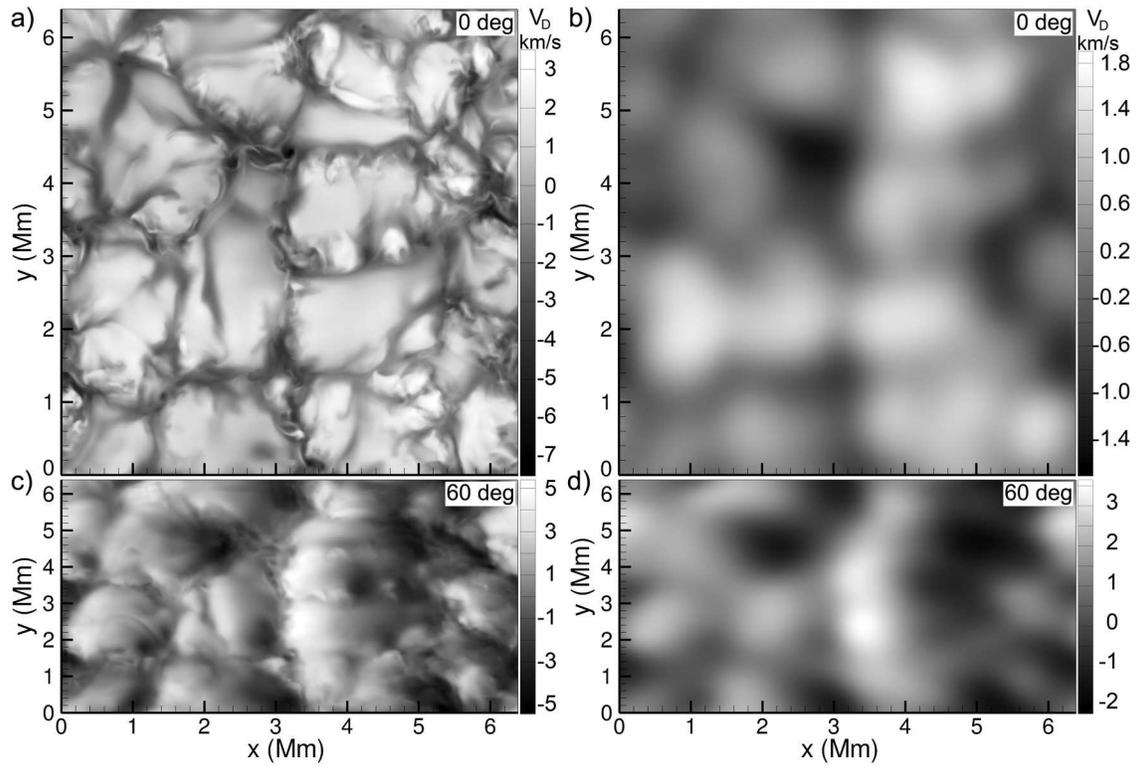}}
 \caption{Doppler-shift snapshots calculated from the simulated profiles of the Fe~I, $\lambda_0 = 6173.3$~\AA~line for the solar disc center, and for $60^0$ from the disc center, for models with the full numerical resolution (12.5~km, panels $a$ and $c$), and with the HMI resolution ($\sim 315$~km/pixel, panels $b$ and $d$).} \label{Doppler}
\end{figure}

\begin{figure}
\centerline{\includegraphics[scale=1]{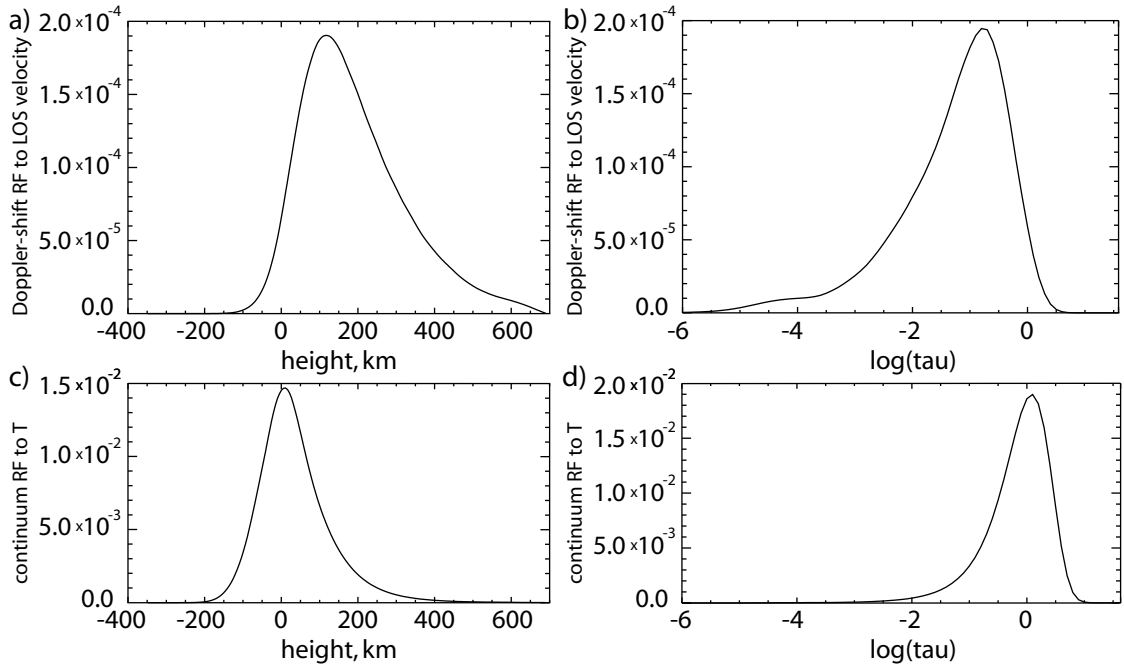}}
 \caption{Response functions (RF) of the HMI Doppler shift observables to line-of-sight (LOS) velocity perturbations  as a function of: $a$) height, $b$) optical depth; and for the HMI continuum intensity to temperature (T) perturbations as a function of: $c$) height, $d$) optical depth.} \label{RF}
\end{figure}

\begin{figure}
 \centerline{\includegraphics[scale=0.8]{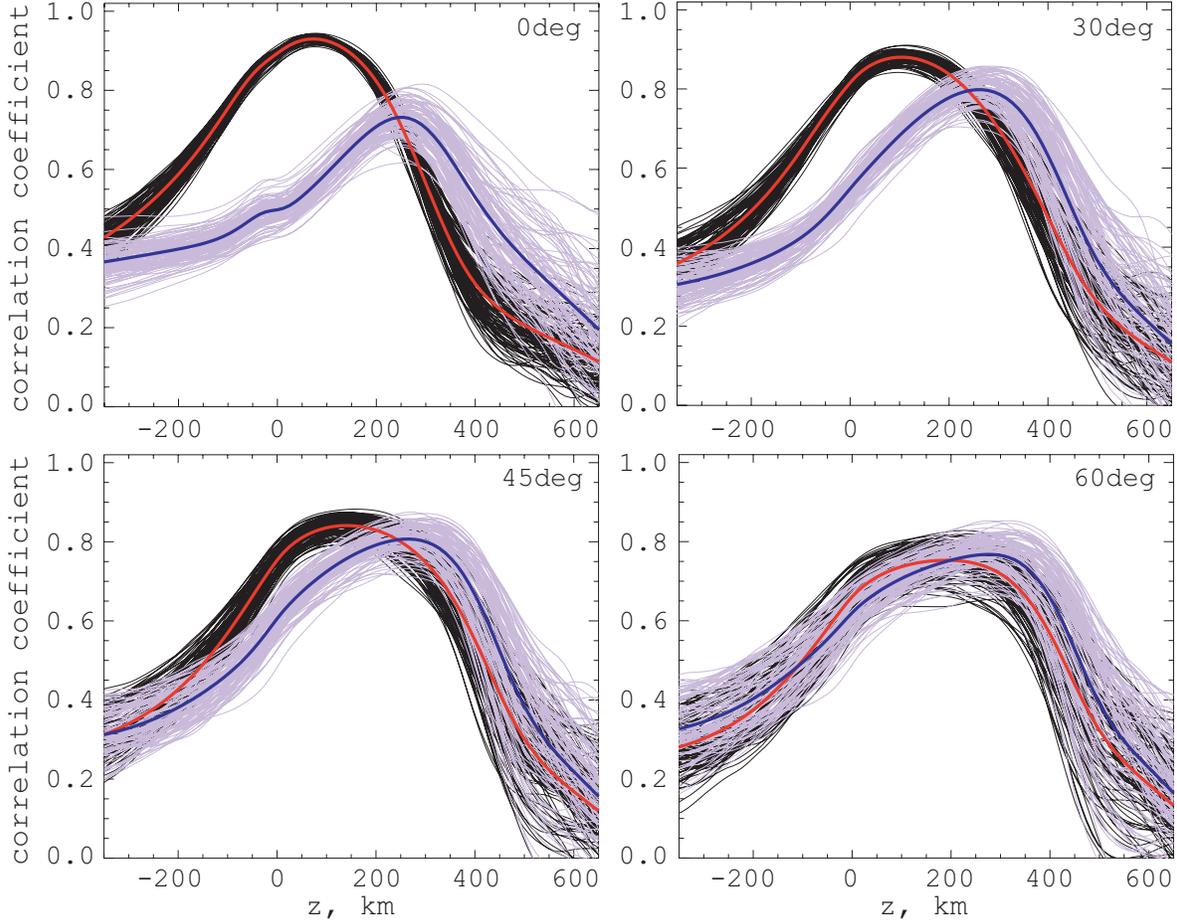}}
 \caption{Correlation coefficients between the Doppler shift and the simulated plasma velocity along the line-of-sight as a function of height in the solar atmosphere for four angular distances from the disc center: $0^0$ (panel $a$), $30^0$ ($b$), $45^0$ ($c$) and $60^0$ (panel $d$). Thin black and pink curves show the correlation coefficient for individual snapshots plotted at 5-min intervals. Black curves correspond to the full numerical resolution data set, and the pink curves correspond to the HMI resolution (modeled by applying the HMI PSF). Thick curves show the mean coefficients averaged over 8 hours for the full resolution data (red curves) and the HMI-resolution data (blue curves).}
 \label{corr-coeff}
 \end{figure}

\begin{figure}
 \centerline{\includegraphics[scale=0.8]{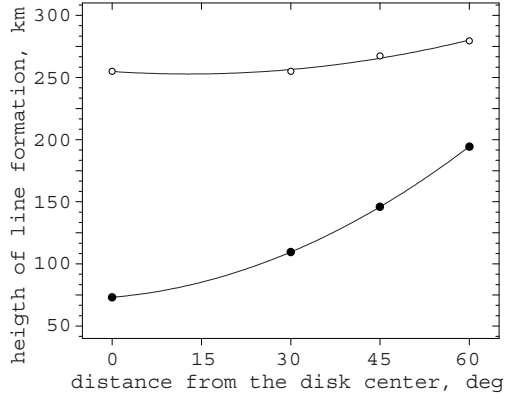}} 
 \caption{Variations of the effective HMI-line formation height for Doppler shift observations as a function of the angular distance from the solar disc center for full numerical resolution (black dots) and HMI resolution (circles). Both sets of data are fitted with second-order polynomials.}
 \label{form}
 \end{figure}

\begin{figure}
 \centerline{\includegraphics[scale=0.8]{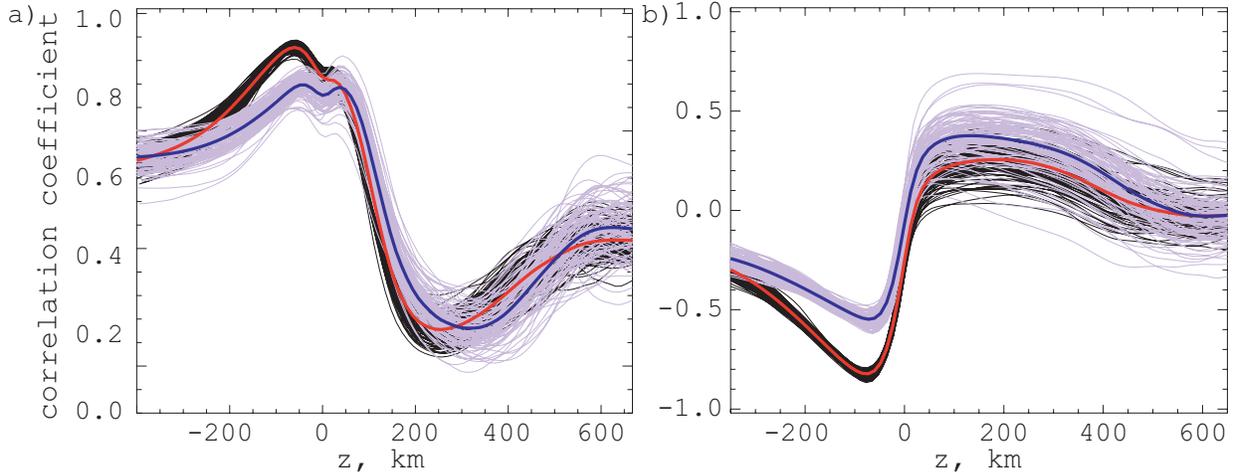}}
 \caption{Correlation coefficients between the continuum intensity and temperature (panel $a$) and density (panel $b$) as a function of height in the solar atmosphere for the disc center. Thin black and pink curves show the correlation coefficient for individual snapshots plotted at 5-min intervals. Black curves correspond to the full numerical resolution data set, and the pink curves correspond to the HMI resolution (modeled by applying the HMI PSF). Thick curves show the mean coefficients averaged over 8 hours for the full resolution data (red curves) and the HMI-resolution data (blue curves).}
 \label{corr-coeff_Ic-T}
 \end{figure}

We found a strong center-to-limb dependence of the line formation height. 
According to the cross-correlation analysis, the maximum correlation of the line-of-sight velocity to the Doppler-shift velocities correspond to a height of about 75~km above the photospheric level ($\tau_{5000}=1$) at the disc center, whereas at $60^0$ from the center the effective formation height increases to about 200~km. 
Note that, for larger angular distances, the estimate of the formation height is more uncertain due to more or less similar correlation between Doppler-shift velocities and actual LOS projected velocities across a wider range of heights. This happens because the observed light crosses a thicker atmospheric layer and experiences a stronger influence of horizontal flows smearing perturbations along the LOS. Such an effect is weaker for the HMI-resolution synthetic data, because most of the small-scale dynamics is not resolved. For the HMI-resolution the center-to-limb variations of the line formation height are smaller: they change from 255~km at the disc center to 275~km at $60^0$. 
Our estimates are somewhat different from the previous results by \cite{Fleck2011} who found in their model a height formation of $\sim 100$ km above the photosphere level for the full resolution data and about 140 km for the data corresponding to the HMI resolution. Our calculations give $\sim 75$~km and 255 km respectively. These discrepancies  may be due to a shallow simulation domain used in the previous work (which was 1.4 Mm in depth and 0.8 Mm in height, while our model is 5.2 Mm in depth and 1 Mm in height). Perhaps, the difference in the depth of the simulated convection zones can lead to differences in the atmospheric structure and dynamics between our and Fleck's simulations, and thus be partially responsible for the difference in the estimate of the height of observed HMI Doppler velocities.
Also, the discrepancies may be due to the adjustment of atomic parameters, such as the oscillator strength and van der Waals coefficient adopted by \cite{Fleck2011} to match the observed line profile. In our simulations we obtain a good agreement with the observational data from the FTS solar atlas  \citep{Brault1987} without any adjustments of the atomic parameters  (Fig.~\ref{line_prof}). 

Also, an effective height of formation of the HMI observables was previously studied by \cite{Wachter2008}  who used the Response Function method  \citep{RuizCobo1994}. These calculations showed that the response function for the HMI-resolution Dopplergrams has a maximum at a height of about 175 km, however, it is extended into higher layers up to $\sim 350$ km. Our height estimate ($\sim 255$ km) is in agreement with the Response Function method. However, we note that this method calculates the sensitivity of the observables to infinitely small variations of physical parameters of a static model of the solar atmosphere, while the cross-correlation approach, used in our paper, calculates the sensitivity to the velocity field of the 3D dynamical model of the solar atmosphere.

\begin{figure}
 \centerline{\includegraphics[scale=1]{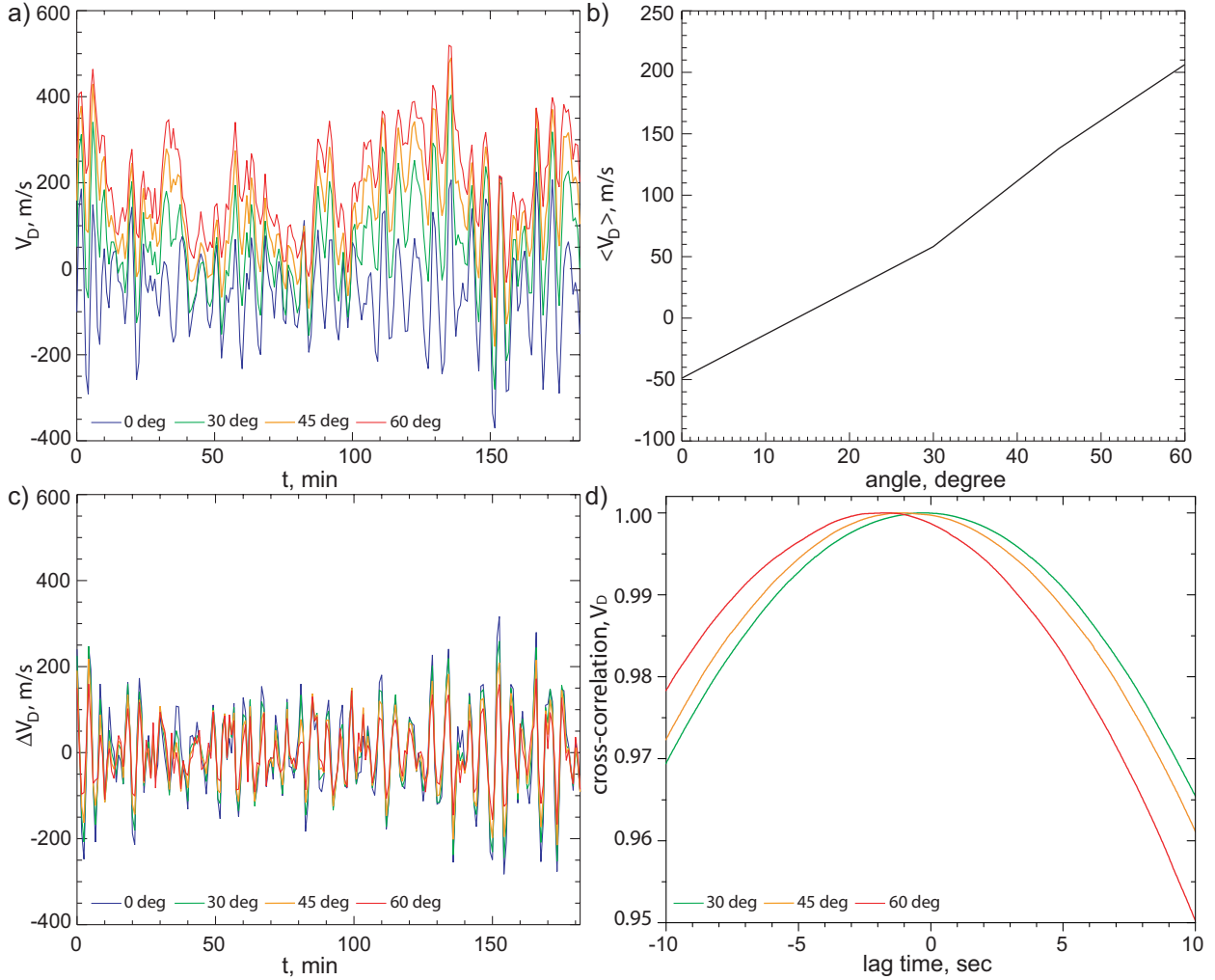}}
 \caption{Panel $a$: Doppler-shift velocity variations as a function of time. Panel $b$: time-averaged Doppler shift velocity as a function of the viewing angle (`blue-shift' effect). Panel $c$: Doppler-shift velocity fluctuations after removing the `blue-shift' effect, showing 5-min oscillations observed for different distances from the disc center. Panel $d$: cross-correlations of the simulated oscillations observed in the Doppler shift at three different angles relative to the disc center.}  \label{velocity_angles_vs_time}
 \end{figure}

\begin{figure}
 \centerline{\includegraphics[scale=0.9]{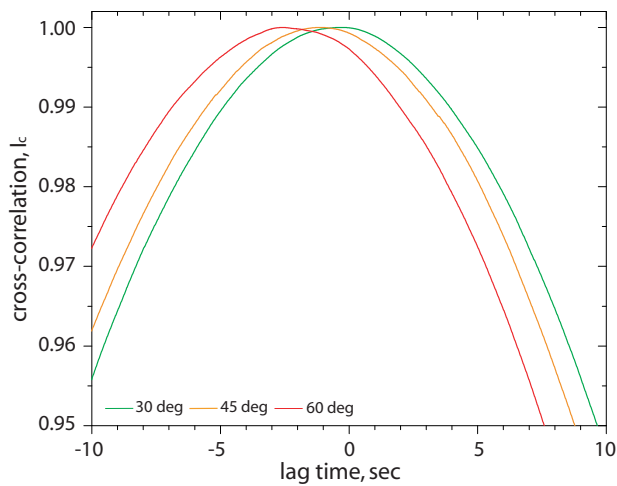}}
 \caption{Cross-correlation functions of the simulated continuum intensity oscillations observed at three different angles $30^0$~(green curve), $45^0$~(orange) and $60^0$~(red curve) relative to the disc center.}
 \label{cross-corr}
 \end{figure}

\begin{figure}
 \centerline{\includegraphics[scale=0.9]{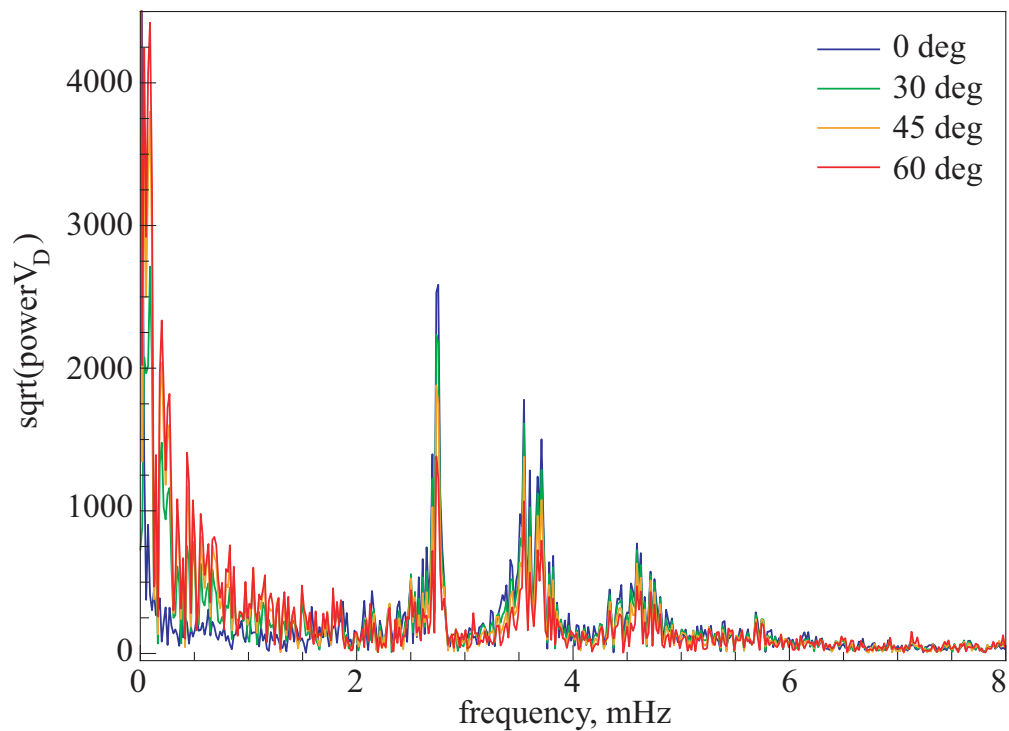}}
 \caption{Square-root of the power spectral density of the simulated Doppler-shift signal for the viewing angles $0^0$~(blue), $30^0$~(green), $45^0$~(orange) and $60^0$~(red). The spectra show the line asymmetry, which is important for helioseismology and ring-diagram analyses.}
 \label{spectrum_velocity_angles}
 \end{figure}

Comparison of the mean Doppler shift for different distances from the disc center reveals a mean blue shift of about 50 -- 100~m/s at the disc center, and a 100 -- 210~m/s red-shift at the $60^0$ distance (Fig.~\ref{velocity_angles_vs_time}~$a$, $b$). In the simulation the mean velocity value fluctuates because of the 5-min oscillations. Removing the systematic mean shifts of the Doppler velocities reveals the decreasing amplitude of the 5-min oscillation for areas closer to the limb (Fig.~\ref{velocity_angles_vs_time}~$c$). Cross-correlation of the Doppler shifts between the same oscillation signal observed at different angles reproduces the helioseismology artifact -- the `concave Sun' effect \citep{Duvall2009,Zhao2012}, observed when the solar oscillation signal from the near-limb regions arrives faster than expected.  This effect is stronger for the continuum intensity data, as is also found in the HMI data. Our results indicate that the systematic travel time shifts found in helioseismology analysis are caused by real physical effects in the solar photosphere. Further modeling work is need to quantify the line formation effects in helioseismology measurements. 

{\it\bf  Acknowledgements.} The simulation results were obtained on the NASA's Pleiades supercomputer at the NASA Ames Research Center. This work was partially supported by the NASA grants NNX10AC55G, NNH11ZDA001N-LWSCSW, NNH13AV81I and Oak Ridge Associated Universities. Also we would like to acknowledge the NASA Ames's user support team and, in particular, Yan-Tyng Chang, Johnny Chang and Steve Heistand.


\begin{thebibliography}{}
\expandafter\ifx\csname natexlab\endcsname\relax\def\natexlab#1{#1}\fi

\bibitem[{{Adam} {et~al.}(1976){Adam}, {Ibbetson}, \& {Petford}}]{Adam1976}
{Adam}, M.~G., {Ibbetson}, P.~A., \& {Petford}, A.~D. 1976, \mnras, 177, 687

\bibitem[{{Baldner} \& {Schou}(2012)}]{Baldner2012}
{Baldner}, C.~S., \& {Schou}, J. 2012, \apjl, 760, L1

\bibitem[{{Bellot Rubio}(2009)}]{BellotRubio2009}
{Bellot Rubio}, L.~R. 2009, \apj, 700, 284

\bibitem[{{Bonet} {et~al.}(2012){Bonet}, {Cabello}, \& {S{\'a}nchez
  Almeida}}]{Bonet2012}
{Bonet}, J.~A., {Cabello}, I., \& {S{\'a}nchez Almeida}, J. 2012, \aap, 539, A6

\bibitem[{{Brault} \& {Neckel}(1987)}]{Brault1987}
{Brault}, W., \& {Neckel}, M. 1987, \solphys, 184, 421

\bibitem[{{Carlsson} {et~al.}(2004){Carlsson}, {Stein}, {Nordlund}, \&
  {Scharmer}}]{Carlsson2004}
{Carlsson}, M., {Stein}, R.~F., {Nordlund}, {\AA}., \& {Scharmer}, G.~B. 2004,
  \apjl, 610, L137

\bibitem[{{Christensen-Dalsgaard}(2004)}]{Christensen-Dalsgaard2004}
{Christensen-Dalsgaard}, J. 2004, in American Institute of Physics Conference
  Series, Vol. 731, Equation-of-State and Phase-Transition in Models of
  Ordinary Astrophysical Matter, ed. V.~{Celebonovic}, D.~{Gough}, \&
  W.~{D{\"a}ppen}, 18--46

\bibitem[{{Christensen-Dalsgaard} {et~al.}(1996){Christensen-Dalsgaard},
  {Dappen}, {Ajukov}, {Anderson}, {Antia}, {Basu}, {Baturin}, {Berthomieu},
  {Chaboyer}, {Chitre}, {Cox}, {Demarque}, {Donatowicz}, {Dziembowski},
  {Gabriel}, {Gough}, {Guenther}, {Guzik}, {Harvey}, {Hill}, {Houdek},
  {Iglesias}, {Kosovichev}, {Leibacher}, {Morel}, {Proffitt}, {Provost},
  {Reiter}, {Rhodes}, {Rogers}, {Roxburgh}, {Thompson}, \&
  {Ulrich}}]{Christensen-Dalsgaard1996}
{Christensen-Dalsgaard}, J., {Dappen}, W., {Ajukov}, S.~V., {et~al.} 1996,
  Science, 272, 1286

\bibitem[{{Couvidat} {et~al.}(2012{\natexlab{a}}){Couvidat}, {Rajaguru},
  {Wachter}, {Sankarasubramanian}, {Schou}, \& {Scherrer}}]{Couvidat2012a}
{Couvidat}, S., {Rajaguru}, S.~P., {Wachter}, R., {et~al.} 2012{\natexlab{a}},
  \solphys, 278, 217

\bibitem[{{Couvidat} {et~al.}(2012{\natexlab{b}}){Couvidat}, {Schou}, {Shine},
  {Bush}, {Miles}, {Scherrer}, \& {Rairden}}]{Couvidat2012}
{Couvidat}, S., {Schou}, J., {Shine}, R.~A., {et~al.} 2012{\natexlab{b}},
  \solphys, 275, 285

\bibitem[{{Duvall} \& {Hanasoge}(2009)}]{Duvall2009}
{Duvall}, Jr., T.~L., \& {Hanasoge}, S.~M. 2009, in Astronomical Society of the
  Pacific Conference Series, Vol. 416, Solar-Stellar Dynamos as Revealed by
  Helio- and Asteroseismology: GONG 2008/SOHO 21, ed. M.~{Dikpati},
  T.~{Arentoft}, I.~{Gonz{\'a}lez Hern{\'a}ndez}, C.~{Lindsey}, \& F.~{Hill},
  103

\bibitem[{{Evershed}(1931)}]{Evershed1931}
{Evershed}, J. 1931, \mnras, 91, 260

\bibitem[{{Feautrier}(1964)}]{Feautrier1964}
{Feautrier}, P. 1964, Comptes Rendus Academie des Sciences (serie non
  specifiee), 258, 3189

\bibitem[{{Fleck} {et~al.}(2011){Fleck}, {Couvidat}, \& {Straus}}]{Fleck2011}
{Fleck}, B., {Couvidat}, S., \& {Straus}, T. 2011, \solphys, 271, 27

\bibitem[{Frutiger(2000)}]{Frutiger2000}
Frutiger, C. 2000, PhD thesis, ETH Zurich

\bibitem[{{Halm}(1907)}]{Halm1907}
{Halm}, J. 1907, Astronomische Nachrichten, 173, 273

\bibitem[{{Hastings}(1873)}]{Hastings1873}
{Hastings}, C.~H. 1873, \nat, 8, 77

\bibitem[{{Jacoutot} {et~al.}(2008){Jacoutot}, {Kosovichev}, {Wray}, \&
  {Mansour}}]{Jacoutot2008}
{Jacoutot}, L., {Kosovichev}, A.~G., {Wray}, A., \& {Mansour}, N.~N. 2008,
  \apjl, 684, L51

\bibitem[{{Kitiashvili} {et~al.}(2013{\natexlab{a}}){Kitiashvili}, {Abramenko},
  {Goode}, {Kosovichev}, {Lele}, {Mansour}, {Wray}, \&
  {Yurchyshyn}}]{Kitiashvili2013b}
{Kitiashvili}, I.~N., {Abramenko}, V.~I., {Goode}, P.~R., {et~al.}
  2013{\natexlab{a}}, Physica Scripta Volume T, 155, 014025

\bibitem[{{Kitiashvili} {et~al.}(2013{\natexlab{b}}){Kitiashvili},
  {Kosovichev}, {Lele}, {Mansour}, \& {Wray}}]{Kitiashvili2013a}
{Kitiashvili}, I.~N., {Kosovichev}, A.~G., {Lele}, S.~K., {Mansour}, N.~N., \&
  {Wray}, A.~A. 2013{\natexlab{b}}, \apj, 770, 37

\bibitem[{{Kosovichev}(2011)}]{Kosovichev2011}
{Kosovichev}, A.~G. 2011, in Lecture Notes in Physics, Berlin Springer Verlag,
  Vol. 832, Lecture Notes in Physics, Berlin Springer Verlag, ed. J.-P.
  {Rozelot} \& C.~{Neiner}, 3--642

\bibitem[{{Nigam} {et~al.}(1998){Nigam}, {Kosovichev}, {Scherrer}, \&
  {Schou}}]{Nigam1998}
{Nigam}, R., {Kosovichev}, A.~G., {Scherrer}, P.~H., \& {Schou}, J. 1998,
  \apjl, 495, L115

\bibitem[{{Norton} {et~al.}(2006){Norton}, {Graham}, {Ulrich}, {Schou},
  {Tomczyk}, {Liu}, {Lites}, {L{\'o}pez Ariste}, {Bush}, {Socas-Navarro}, \&
  {Scherrer}}]{Norton2006}
{Norton}, A.~A., {Graham}, J.~P., {Ulrich}, R.~K., {et~al.} 2006, \solphys,
  239, 69

\bibitem[{{Ruiz Cobo} \& {del Toro Iniesta}(1994)}]{RuizCobo1994}
{Ruiz Cobo}, B., \& {del Toro Iniesta}, J.~C. 1994, \aap, 283, 129

\bibitem[{{Scherrer} {et~al.}(1995){Scherrer}, {Bogart}, {Bush}, {Hoeksema},
  {Kosovichev}, {Schou}, {Rosenberg}, {Springer}, {Tarbell}, {Title},
  {Wolfson}, {Zayer}, \& {MDI Engineering Team}}]{Scherrer1995a}
{Scherrer}, P.~H., {Bogart}, R.~S., {Bush}, R.~I., {et~al.} 1995, \solphys,
  162, 129

\bibitem[{{Scherrer} {et~al.}(2012){Scherrer}, {Schou}, {Bush}, {Kosovichev},
  {Bogart}, {Hoeksema}, {Liu}, {Duvall}, {Zhao}, {Title}, {Schrijver},
  {Tarbell}, \& {Tomczyk}}]{Scherrer2012}
{Scherrer}, P.~H., {Schou}, J., {Bush}, R.~I., {et~al.} 2012, \solphys, 275,
  207

\bibitem[{{Schou} {et~al.}(2012){Schou}, {Scherrer}, {Bush}, {Wachter},
  {Couvidat}, {Rabello-Soares}, {Bogart}, {Hoeksema}, {Liu}, {Duvall}, {Akin},
  {Allard}, {Miles}, {Rairden}, {Shine}, {Tarbell}, {Title}, {Wolfson},
  {Elmore}, {Norton}, \& {Tomczyk}}]{Schou2012}
{Schou}, J., {Scherrer}, P.~H., {Bush}, R.~I., {et~al.} 2012, \solphys, 275,
  229

\bibitem[{{Solanki} {et~al.}(1992){Solanki}, {Rueedi}, \&
  {Livingston}}]{Solanki1992}
{Solanki}, S.~K., {Rueedi}, I.~K., \& {Livingston}, W. 1992, \aap, 263, 312

\bibitem[{{Stenflo} {et~al.}(1997){Stenflo}, {Bianda}, {Keller}, \&
  {Solanki}}]{Stenflo1997}
{Stenflo}, J.~O., {Bianda}, M., {Keller}, C.~U., \& {Solanki}, S.~K. 1997,
  \aap, 322, 985

\bibitem[{{van Noort}(2012)}]{vanNoort2012}
{van Noort}, M. 2012, \aap, 548, A5

\bibitem[{{Wachter}(2008)}]{Wachter2008}
{Wachter}, R. 2008, \solphys, 251, 491

\bibitem[{{Wachter} {et~al.}(2012){Wachter}, {Schou}, {Rabello-Soares},
  {Miles}, {Duvall}, \& {Bush}}]{Wachter2012}
{Wachter}, R., {Schou}, J., {Rabello-Soares}, M.~C., {et~al.} 2012, \solphys,
  275, 261

\bibitem[{{Zhao} {et~al.}(2012{\natexlab{a}}){Zhao}, {Nagashima}, {Bogart},
  {Kosovichev}, \& {Duvall}}]{Zhao2012}
{Zhao}, J., {Nagashima}, K., {Bogart}, R.~S., {Kosovichev}, A.~G., \& {Duvall},
  Jr., T.~L. 2012{\natexlab{a}}, \apjl, 749, L5

\bibitem[{{Zhao} {et~al.}(2012{\natexlab{b}}){Zhao}, {Couvidat}, {Bogart},
  {Parchevsky}, {Birch}, {Duvall}, {Beck}, {Kosovichev}, \&
  {Scherrer}}]{Zhao2012a}
{Zhao}, J., {Couvidat}, S., {Bogart}, R.~S., {et~al.} 2012{\natexlab{b}},
  \solphys, 275, 375

\end{thebibliography}

\end{document}